\documentclass[11pt]{article}
\usepackage{authblk}
\usepackage{paralist}
\usepackage{amsmath}
\usepackage{graphicx}
\usepackage{hyperref}
\usepackage[font=footnotesize]{caption}
\usepackage[numbers]{natbib}

\begin{document}
\title{Pulsar Acceleration Searches on the GPU for the Square Kilometre Array}
\author[1]{Sofia~Dimoudi}
\author[1]{Wesley~Armour}
\date{}
\affil[1]{University of Oxford, Oxford, United Kingdom \\\texttt{sofia.dimoudi@oerc.ox.ac.uk}}
\maketitle
\begin{abstract}
Pulsar acceleration searches are methods for recovering signals from radio telescopes, that may otherwise be lost due to the effect of orbital acceleration in binary systems. The vast amount of data that will be produced by next generation instruments such as the Square Kilometre Array (SKA) necessitates real-time acceleration searches, which in turn requires the use of HPC platforms. We present our implementation of the Fourier Domain Acceleration Search (FDAS) algorithm on Graphics Processor Units (GPUs) in the context of the SKA, as part of the Astro-Accelerate real-time data processing library, currently under development at the Oxford e-Research Centre (OeRC), University of Oxford.
\end{abstract}
\section{Introduction}
Pulsars are highly magnetized rotating neutron stars which emit a beam of electromagnetic radiation from their magnetic poles. When the beam passes our line of sight it is observed as a pulse of radiation. The period of the pulses is very precise and extremely regular, a property that makes these objects interesting for a variety of astronomical studies. Pulsars in binary systems which are found typically with millisecond pulse periods, are particularly interesting as they enable very high precision measurements which can be used to test theories of gravity \citep{Lorimer-2005}.

The standard way to identify pulsar signals in a radio telescope signal is to run a periodicity search by applying the Discrete Fourier Transform (DFT) on the incoming time series data. The power spectrum produced by the DFT can reveal periodic signal power concentrated in the frequencies of the received pulse harmonics. In the case of binary pulsars however, due to the Doppler effect, orbital motion creates a drift in the apparent frequency as the object is moving towards and away from us. As a result, the recovered power spreads over neighbouring frequencies, which can have a significant effect on the amount of power that can be recovered with this method, and can potentially prevent detection. In order to correct for this effect,  a simple model is used which assumes a constant acceleration for a small fraction of the orbital period. A search over a number of trials for different constant acceleration values can then show increased detection at trials corresponding to an accelerated pulsar. This type of search is called an acceleration search. Acceleration search techniques have by now been well studied (e.g. in \citep{Middleditch-1986, Anderson-1990, Johnston-1991, Ransom-2002}), and currently there are two main methods that are commonly used. The first one works in the time domain by resampling the time series according to a time offset caused by the constant acceleration, followed by a standard periodicity search for each acceleration value. This technique can be computationally very expensive, as it requires the calculation of many long DFTs. A different approach is to create a set of Finite Impulse Response (FIR) filters in the Fourier domain that describe the effect of constant acceleration, and then correlate each one with the signal. This is the correlation technique \citep{Ransom-2002} and can be much more efficient because the filter size can be small and the computations can be performed in parallel, as many short independent Fourier transforms and convolutions. 

Acceleration searches are critical in fulfilling some of the key scientific objectives for the next generation radio telescope, the Square Kilometre Array (SKA). The SKA will use pulsars to search for gravitational waves, and test general relativity under very strong-field gravity conditions. Prior to conducting any periodicity search, the signals need to be analysed for determining their Dispersion Measure (DM). This process, known as de-dispersion will result in a large number of data series that will need to be searched individually for the existence of periodic signals. The total processing must be done in real-time, which is limited to the observation period. We work with the assumption that a time series produced from an SKA beam will contain approximately 8 million samples, the de-dispersion will produce \(\scriptstyle \sim \)\,6000 DMs and the observation period will be approximately 530 seconds. These restrictions, along with the vast data volume produced by the SKA, increase the demand for fast algorithms and for the use of energy efficient HPC.  A real-time time-domain data processing library, Astro-Accelerate \citep{Armour:2011vw}, is currently being developed at the Oxford e-Research centre as part of the time domain science efforts for the SKA. Astro-Accelerate will be using specialised algorithms, targeted in many-core accelerators, with the aim of fulfilling these limitations, and has been successfully used in \citep{karastergiou2015limits}. 

This document presents an overview of the implementation of the FDAS on GPUs that will form part of this work.
\section{Fourier Domain Acceleration Search Method}
The FDAS method was proposed by \citet{Ransom-2002} who devised a correlation technique and derived the mathematical form of a finite impulse response filter template in the Fourier space that models the effect of constant acceleration. By applying the inverse filter, that is, the complex conjugate of the filter to the Fourier transformed series, it is shown that the signal power that has leaked to surrounding Fourier bins is recovered in a single frequency bin. The corrected signal response resulting from the correlation technique is expressed by the relationship
\begin{equation}
\mathcal{F}_{r_0} = \sum_{k=r_0-m/2}^{k=r_0+m/2} \mathcal{F}_k \mathcal{F}_{r_0-k}^{\ast} \,,
\end{equation}
where $\mathcal{F}_{r_0-k}^{\ast}$ is the inverse filter template and $\mathcal{F}_k$ the signal Fourier response at bin $k$, with a frequency offset $|r_0 - k|$ from a reference frequency $r_0$.

\begin{figure}[htbp]
\begin{center}
\includegraphics[width=0.7\textwidth]{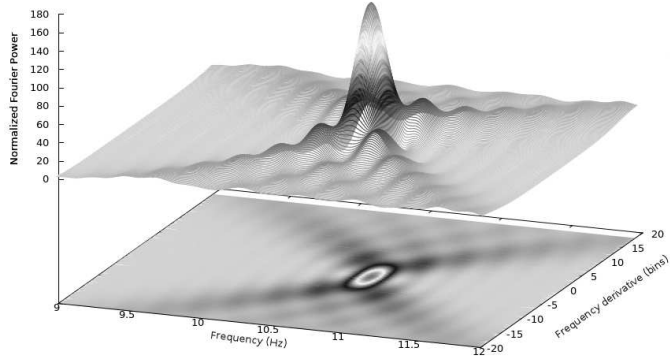}
\caption{Example of  a theoretical one harmonic with orbital acceleration on an $f - \dot{f}$ plane.}
\label{fig:fig1}
\end{center}
\end{figure}

Each template is related to orbital acceleration via a frequency derivative $\dot{f}$ which expresses the frequency drift due to the Doppler effect corresponding to the constant acceleration, and is directly proportional to the number of frequency bins the signal would spread into, i.e.\  $\dot{r}=\dot{f}T^2$, where $T^2$ is the observation period.
To examine the signal for a range of orbital acceleration values, multiple templates must be used. The result from this processing can be visualised on a two-dimensional $f - \dot{f}$ plane which reveals the position of the signal Fourier power in frequency and orbital acceleration. An example of this plane is shown in figure~\ref{fig:fig1}.

Technically, the correlation of the signal with multiple templates is a matched filtering process, and can be done efficiently in the Fourier domain using the convolution theorem with an overlap-save method \citep{press1992num}.

\section{High Performance Computing}
In the last decade there has been an increasing interest in using co-processors with parallel architectures as hardware accelerators, such as Graphical Processing Units (GPU), Field Programmable Gate Arrays (FPGA), and more recently the Intel Xeon Phi. NVIDIA GPUs are now well established as high performance devices. They have a rich software ecosystem, a proven track record, and are one of the highest performing co-processors. 

The FDAS method as described earlier has a significant part devoted to applying DFTs, and as such, we use the proven CUFFT library \citep{cufft-2015} for this task. The problem structure is also suitable as the data are arranged in small mostly independent blocks which fits the GPU programming model. Trends towards increasing the GPU performance per Watt ratio make NVIDIA GPUs a viable solution for the FDAS component of the SKA signal processing pipeline.
\section{Implementation and Preliminary Results}
\begin{figure}[htbp]
\begin{center}
\includegraphics[width=\textwidth]{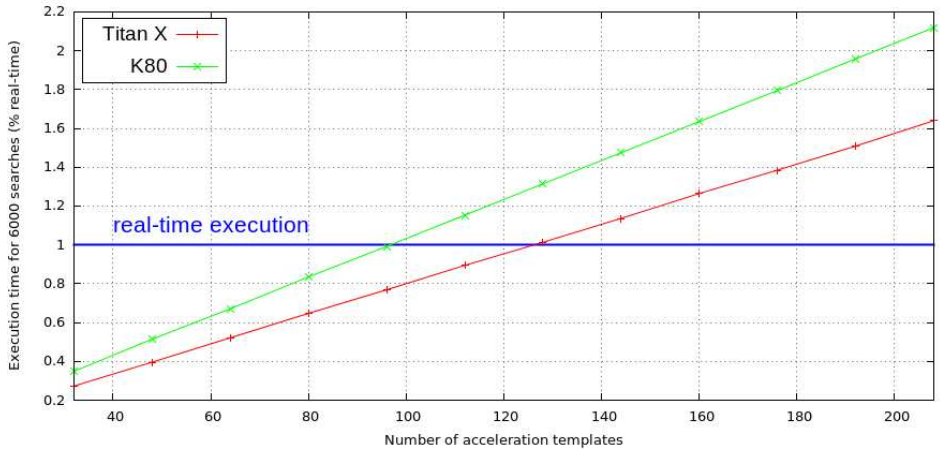}
\caption{Real-time performance of Fourier domain acceleration search on a GeForce Titan X and a Tesla K80 (using a single GPU on the card).}
\label{fig:fig2}
\end{center}
\end{figure}

After an initialization stage that includes computation of the templates to be used, the algorithm performs the following conceptual steps for each received time series:
\begin{inparaenum}
\item The time series is transferred to the GPU via PCI-express, this can be done asynchronously;
\item Real to complex DFT of the time series; \label{itm:rfft}
\item Complex DFT output is divided in segments. Each segment contains a region from the previous segment to correct for edge effects; \label{itm:ovlap}
\item Complex to complex DFT of each segment; \label{itm:cfft}
\item Fourier Domain convolution of each segment with all templates (complex multiply - scale), 2-dimensional $f - \dot{f}$ plane produced; \label{itm:convol}
\item Inverse complex to complex DFT of each $f - \dot{f}$ segment; and \label{itm:cifft}
\item Power spectrum operations.\label{itm:ffdotpow}
\end{inparaenum} 
Steps~\ref{itm:rfft},~\ref{itm:cfft} and~\ref{itm:cifft} were done using the CUFFT library. Step~\ref{itm:ovlap} is a simple copy of overlapped data segments to an extended array so that they are independent and contiguous, and has a negligible cost. The actual filtering operations are performed as Fourier domain convolutions via complex element-wise multiplications in step~\ref{itm:convol}. This step uses a custom GPU kernel which has been optimized for memory transfers and data reuse. The power spectrum operations in step~\ref{itm:ffdotpow} currently consists of the main calculation of Fourier powers and edge removal of the convolved segments. 
We performed tests on time series containing $2^{23}$ samples with over 100 templates, which we believe is consistent with the current SKA specification. The tests were run on two NVIDIA cards: a GeForce Titan X of the newest Maxwell architecture, and a dual-GPU Tesla K80, using only one of its 2 GPUs. Figure~\ref{fig:fig2} illustrates the real-time performance of the algorithm relative to the number of acceleration templates applied.  As can be seen, at this stage, we are able to achieve better than real-time performance for the SKA requirement. 
%\articlefigure[width=0.8\textwidth]{P031_f2.eps}{fig2}{Real-time performance of Fourier domain acceleration search on a GeForce Titan X and a Tesla K80 (using a single GPU on the card).}
\section*{Acknowledgements}
This work is supported by a Leverhulme Trust Project Grant (ARTEMIS: Real-time discovery in Radio Astronomy). It has also received support from members of the Oxford pulsar group, namely Aris Karastergiou, Christopher Williams and Jayanth Chennamangalam, as well as support from the Time Domain Team, a collaboration between Oxford, Manchester and MPIfR Bonn, to design and build the SKA pulsar search capabilities.
\bibliographystyle{kp}
\bibliography{adass-arXiv}  % For BibTex

\begingroup\raggedright\begin{thebibliography}{9}
\expandafter\ifx\csname natexlab\endcsname\relax\def\natexlab#1{#1}\fi

\bibitem[{Lorimer} and {Kramer}(2004)]{Lorimer-2005}
D.~R. {Lorimer} and M.~{Kramer}, ``{Handbook of Pulsar Astronomy}'', {Cambridge
  University Press}, {Cambridge, UK}, 2004.

\bibitem[{Middleditch} and {Priedhorsky}(1986)]{Middleditch-1986}
J.~{Middleditch} and W.~C. {Priedhorsky}, ``{Discovery of rapid quasi-periodic
  oscillations in Scorpius X-1}'', {\em Astrophysical Journal} {\bfseries 306}
  (1986) 230--237.

\bibitem[Anderson et~al.(1990)Anderson, Gorham, Kulkarni, Prince, and
  Wolszczan]{Anderson-1990}
S.~B. Anderson, P.~W. Gorham, S.~R. Kulkarni, T.~A. Prince, and A.~Wolszczan,
  ``{Discovery of two radio pulsars in the globular cluster M15}'', {\em
  Nature} {\bfseries 346} 07 (1990) 42--44.

\bibitem[Johnston and Kulkarni(1991)]{Johnston-1991}
H.~M. Johnston and S.~R. Kulkarni, ``On the detectability of pulsars in close
  binary systems'', {\em The Astrophysical Journal} {\bfseries 368} (1991)
  504--514.

\bibitem[Ransom et~al.(2002)Ransom, Eikenberry, and Middleditch]{Ransom-2002}
S.~M. Ransom, S.~S. Eikenberry, and J.~Middleditch, ``Fourier techniques for
  very long astrophysical time-series analysis'', {\em The Astronomical
  Journal} {\bfseries 124} (2002), no.~3, 1788.

\bibitem[Armour et~al.(2012)Armour, Karastergiou, Giles, Williams, Magro,
  Zagkouris, Roberts, Salvini, Dulwich, and Mort]{Armour:2011vw}
W.~Armour, A.~Karastergiou, M.~Giles, C.~Williams, A.~Magro, K.~Zagkouris,
  S.~Roberts, S.~Salvini, F.~Dulwich, and B.~Mort, ``{A GPU-based survey for
  millisecond radio transients using ARTEMIS}'', {\em ASP Conf. Ser.}
  {\bfseries 461} (2012) 33--36,
 \href{http://xxx.lanl.gov/abs/1111.6399}{{\ttfamily arXiv:1111.6399}}.
%%CITATION = ARXIV:1111.6399;%%.

\bibitem[Karastergiou et~al.(2015)Karastergiou, Chennamangalam, Armour,
  Williams, Mort, Dulwich, Salvini, Magro, Roberts, Serylak,
  et~al.]{karastergiou2015limits}
A.~Karastergiou, J.~Chennamangalam, W.~Armour, C.~Williams, B.~Mort,
  F.~Dulwich, S.~Salvini, A.~Magro, S.~Roberts, M.~Serylak, {\em et~al.},
  ``{Limits on fast radio bursts at 145 MHz with artemis, a real-time software
  backend}'', {\em Monthly Notices of the Royal Astronomical Society}
  {\bfseries 452} (2015), no.~2, 1254--1262.

\bibitem[Press(1992)]{press1992num}
W.~Press, ``Numerical recipes in c: The art of scientific computing'',
  Cambridge University Press, 1992.

\bibitem[{NVIDIA Corporation}(2015)]{cufft-2015}
{NVIDIA Corporation}, {\em {cuFFT Library User's Guide, v7.5}}, 2015.

\end{thebibliography}\endgroup
\end{document}